\documentclass{article}
\usepackage{spconf,amsmath,epsfig}
\usepackage[inline]{enumitem}
\usepackage{multirow}

\title{FIRE RISK ANALYSIS BY USING SENTINEL-2 DATA: THE CASE STUDY OF THE VESUVIUS IN CAMPANIA, ITALY}
\name{Domenico A. G. Dell'Aglio, Massimiliano Gargiulo, Antonio Iodice, Daniele Riccio, Giuseppe Ruello}
\address{University of Naples Federico II\\
    Department of Electrical Engineering and Information Technology\\
    Via Claudio, 21, 80125, Naples, Italy}
\begin{document}
%
\maketitle
\begin{abstract}
As sadly known, forest fires are part of a set of natural disasters that have always affected regions of the world typically characterized by a tropical climate with long periods of drought. However, due to climate changes of the recent years, other regions of our planet that were not affected by this plague have also had to deal with this phenomenon. One of them is certainly the Italian peninsula, and especially the regions of southern Italy. For this reason, the scientific community, and in particular that one of the remote sensing, plays an important role in the development of reliable techniques to provide useful support to the competent authorities. Therefore, in this work, the capability of the Normalized Differential Water Index (NDWI), derived from spaceborne remote sensing (RS) data, is assessed to monitor the forest fires occurred on a specific study area during the summer of 2017: the volcano Vesuvius, near Naples (in Campania, Italy). In particular, the index is obtained from Sentinel-2 multispectral images of the European Space Agency (ESA), which are free of charge and open accessible. Moreover, the twin Sentinel-2 (S-2) sensors allows to overcome some restrictions on time delivery and high frequency observation. These requirements are goodly matched by other spaceborne sensors, such as MODIS and VIIRS satellites, but at the expense of a lower spatial resolution.
\end{abstract}
\section{INTRODUCTION}
\label{sec:intro}

Nowadays the Earth observations are extensively used to manage risk (floods, subsidence, landslide), to monitor forest, land cover/land use, and so on. For instance, in the forest management the damages caused by fires are a key element both for the environment properties and for human life. In fact, fires are the main cause of forest loss around the world and even animal heritage loss. Therefore, this general impoverishment of the ecosystem corresponds to an increase of the hydro-geological instability, especially in mountainous areas with high slopes, threatening any human settlements downstream and around the burned areas. Although on one the hand, these tragic events have highlighted the limits of the current preventive measures, on the other hand they have determined the need to find alternative ways that can support the existing ones and the relevant authorities. 

\begin{figure}[t]
\begin{minipage}[b]{.48\linewidth}
  \centering
  \centerline{\epsfig{figure=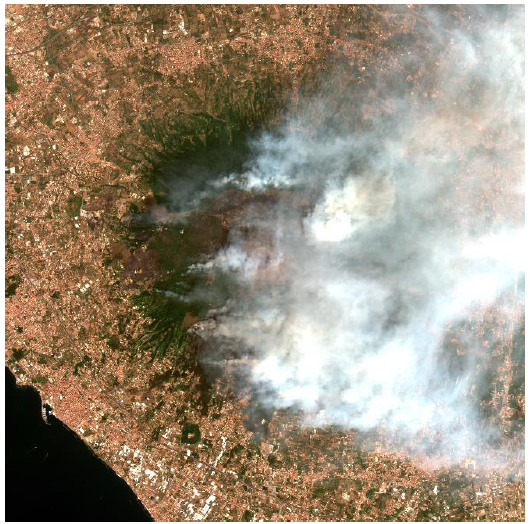,width=4.0cm}}
  \centerline{(a)}\medskip
\end{minipage}
\hfill
\begin{minipage}[b]{0.48\linewidth}
  \centering
  \centerline{\epsfig{figure=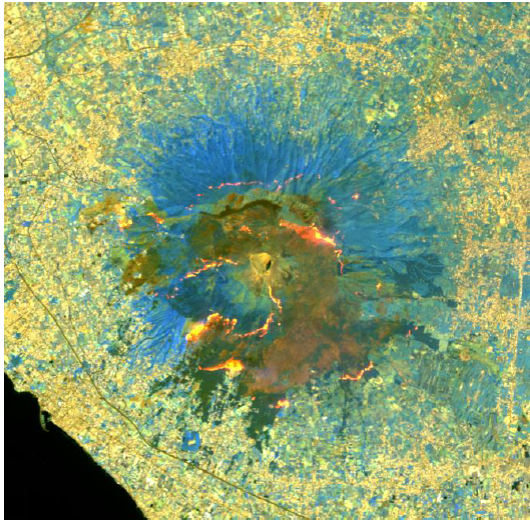,width=4.0cm}}
  \centerline{(b)}\medskip
\end{minipage}
\caption{RGB image of the fire on Vesuvius from S-2 data acquired on 12 July 2017 (a) and false colour composite ($\rho_{12}$, $\rho_{11}$ and $\rho_{8A}$ bands) of the same data (b).}
\label{fig:fig1}
\end{figure}

Therefore, the satellite remote sensing can certainly be helpful to support forest fires monitoring activities. It allows some important advantages such as the ability to find information on large areas in relatively quick way  (including those inaccessible and/or dangerous to humans). At the moment, fires monitoring services based on the processing of satellite data \cite{Giglio98,Giglio2014} have three main aims:
\begin{enumerate*}[label=(\roman*)]
    \item fire risk assessment,
    \item active fire detection, and
    \item burned area mapping.
\end{enumerate*}
Among these, the first is certainly the most difficult to handle for the remote sensing scientists. In fact, to the best of our knowledge, a specific technique able to well perform this task is still absent. Nowadays, prevention mechanisms are mainly based on the processing of meteorological data and on the use of human resources. Therefore, in order to find a relationship between the fire ignition event and the status of the vegetation, a time series based on Normalized Difference Water Index (NDWI) \cite{Gao} products, achieved by S-2 data, has been used in this work. Moreover, even vegetation indices such as the Normalized Difference Vegetation Index (NDVI) are commonly considered into fire risk assessment \cite{yu2017fire}. 

The Fire Risk monitoring is very important, considering that about 75.000 hectares of woodland were destroyed in Italy, and in particular 65\% of the total damages in the southern regions, in 2017. For this analysis, we selected some examples of fires that devastated the Campania region, in south of Italy, during the summer of 2017. Among these burned areas important forest heritage was also involved, such as that ones of the volcano Vesuvius (Fig.\ref{fig:fig1}), which includes a National Park of about 7.259 hectares.

In this paper we consider an analysis of the fire risk prevention in this area. Firstly, in the part \ref{subsec:dataset} of section \ref{sec:method} we described in more details the dataset and the investigated study area. Then, in the part \ref{subsec:NDWI_trend} of section \ref{sec:method} we focused on the proposed multi-temporal analysis and some results are provided in section \ref{sec:results}. Eventually, we highlighted on the conclusions and the future works.

\section{PROPOSED ANALYSIS}
\label{sec:method}

\newcommand{\ru}{\rule{0mm}{3mm}}
\begin{table}[!b]
\centering
\def\arraystretch{1.25}
\setlength{\tabcolsep}{4.8pt}
\footnotesize
\begin{tabular}{l|c|c}  
\hline
\ru Spatial  & Bands & Wavelength Range \\
\ru Resolution [$m$] & (Bands Number) & [${\mu}m$]\\
\hline
\ru 10 & Blue (2), Green (3), Red (4), & \multirow{2}{*}{$0.490 - 0.842$ }\\
\ru & and NIR (8)& \\
\hline
\ru 20 & Vegetation Red Edge (5,6,7, 8A) &\multirow{2}{*}{$0.705 - 2.190$} \\
\ru & and SWIR (11,12) & \\
\hline
\ru 60 & Coastal Aerosol (1), & \multirow{2}{*}{$0.443 - 1.375$ }\\
\ru & Water Vapour (9), and SWIR (10) & \\
\hline
\end{tabular}
\caption{\small The 13 Sentinel-2 bands.}
\label{tab:tab2}
\end{table}

In order to detect wildfires and monitoring forest areas under fire risk, current remote sensing systems, such as the European Forest Fire Information System (EFFIS) \cite{san2002towards} or the Advanced Fire Information System (AFIS) \cite{mcferren2009southern}, are essentially based on data processing from MODIS \cite{Giglio98} and VIIRS \cite{Giglio2014} satellite sensors. On the one hand, they return multispectral images that have good performance in terms of frequency of observations (16 days for both sensors). On the other hand, they have a very low spatial resolution on the ground (that is about hundreds of meters), resulting in an inability to detect the exact position of active fires and the exact extension of the burned areas.

\subsection{Dataset and Study Area}
\label{subsec:dataset}

The above-mentioned sensors are not able to analyse the considered are because of its small dimension and of the wildfires extension. The small fires aftermaths are surely less significant than large fires in terms of climate changes, but play a significant role in the usage of land and in total gas emissions. In order to overcome these bottlenecks, some works have used data from the Landsat and the ESA Sentinel-2A/2B \cite{drusch2012sentinel,immitzer2016first,cicala2018landsat} satellite sensors.

The Landsat bands, from Visible to Short Wave InfraRed (SWIR), are provided at 30-m spatial resolution whereas the Panchromatic at 15-m, on the contrary these bands from Sentinel-2 are acquired at different spatial resolution, from 60-m to 10-m, and we can see the details in Tab.\ref{tab:tab2}. Moreover, the S-2 data even provide excellent performance in terms of acquisition rate (passages on same area every 5 days through the combined use the twin satellite sensors). As we can see in Tab.\ref{tab:tab2}, they can acquire data on 13 spectral channels in the Visible, Near InfraRed (NIR) and SWIR bands at different spatial resolution \cite{drusch2012sentinel,immitzer2016first}. However, S-2 NIR and SWIR bands are provided at 10-m and 20-m spatial resolution respectively, as reported in Tab.\ref{tab:tab2}. In order to further improve the spatial resolution and to obtain more accurate fires mapping, some works resort to super resolution techniques on SWIR bands. Thus, the 10-m spatial resolution for SWIR bands is achieved using data from the S-2 bands at higher resolution \cite{dell2019active,gargiulo1906cnn}.

\begin{table}[!b]
\centering
\def\arraystretch{1.25}
\setlength{\tabcolsep}{9.2pt}
\footnotesize
\begin{tabular}{ccccc}
\hline
\hline
\textbf{Acquisition date} & \textbf{Mission} & \textbf{Cloud cover} & \textbf{Fire event} \\
\hline
\hline
09/12/2016 & S-2A & NO & - \\
29/03/2017 & S-2A & NO & - \\
08/04/2017 & S-2A & NO & - \\
27/06/2017 & S-2A & PARTIALLY & pre-fire \\
07/07/2017 & S-2A & PARTIALLY & post-fire \\
12/07/2017 & S-2B & NO & during-fire \\
17/07/2017 & S-2A & PARTIALLY & during-fire \\
06/08/2017 & S-2A & NO & post-fire \\
16/08/2017 & S-2A & NO & - \\
31/08/2017 & S-2B & NO & - \\
15/10/2017 & S-2A & NO & - \\
04/11/2017 & S-2A & NO & - \\
19/12/2017 & S-2B & NO & - \\
24/12/2017 & S-2A & NO & - \\
\hline
\hline
\end{tabular}
\caption{\small Acquisition dataset from Sentinel-2 mission used for this study.}
\label{tab:tab1}
\end{table}

For the specific investigated study area, we have used The Level-2A products. The L2A products represents the Bottom-Of-Atmosphere (BOA) reflectances that means atmospheric corrected reflectance images. The BOA reflectances are required when we evaluate a multi-temporal analysis based on spectral index values. 
In particular, we select several patches from the volcano Vesuvius zone (taking into account the absence of cloud cover). In Tab. \ref{tab:tab1} we can see the whole dataset exploited for the proposed analysis, where the presence or not of the cloud-covering is also shown.

\subsection{Multi-temporal NDWI trend}
\label{subsec:NDWI_trend}

\begin{figure}[!t]
\begin{minipage}[b]{1.0\linewidth}
  \centering
  \centerline{\epsfig{figure=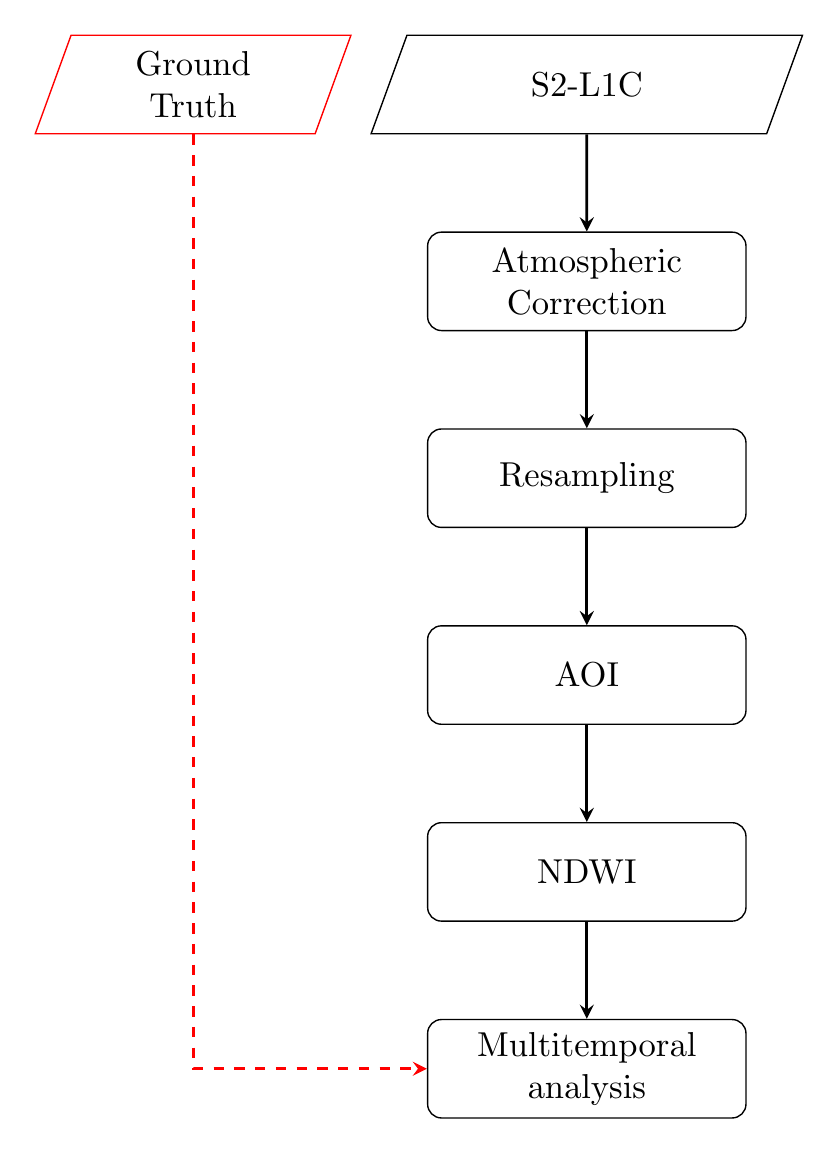,width=4.5cm}}
\end{minipage}
\caption{General workflow of the proposed analysis.}
\label{fig:flow}
\end{figure}

In this section we present the analysis conducted on the aforementioned target site for the proposed topic. The general workflow is depicted in Fig.\ref{fig:flow}.
Moving downwards, we start from the L1C products of Sentinel-2 that correspond to the Top-Of-Atmosphere (TOA) reflectances, and on these reflectances we consider the atmospheric correction. Then, we consider a resampling of all the S-2 bands, in fact we bring them back to 20-m, and we limit the entire dataset to the study area (AOI).  Eventually, we compute the NDWI 
On Sentinel-2 data the NDWI is defined as \cite{Gao}:

\begin{equation} \label{eq:eq1}
    NDWI = \frac{\rho_{8} - \rho_{11}}{\rho_{8} + \rho_{11}}
\end{equation}

\noindent where $\rho_{8}$ is the NIR band, centered at the wavelength of 842 \textit{nm}; while $\rho_{11}$ is the SWIR band, centered at 1610 \textit{nm}. Because of its main characteristic to monitor changes in water content of leaves \cite{Gao}, the NDWI has been used to exploit a time series analysis from Sentinel-2 Level-2A data covering almost one year of acquisitions (from December 2016 to December 2017), see Tab.\ref{tab:tab1}. 
Furthermore, because of densely populated area, in this multi-temporal NDWI-based analysis both dry vegetation and artificial surfaces can be found. In order to remove the pixels related to the artificial surfaces and ensure to only evaluate the time series on vegetation pixels, we consider a further threshold on the NDVI, a most popular vegetation index, defined as in \cite{yu2017fire}:

\begin{equation} \label{eq:eq1}
    NDVI = \frac{\rho_{8} - \rho_{4}}{\rho_{8} + \rho_{4}}
\end{equation}
\noindent where $\rho_{8}$ is the above-mentioned NIR band, and  $\rho_{4}$ is the Red band, centered at 665 \textit{nm}.

Therefore, for each selected Area Of Interest (AOI) and for all the data frame, the average values of NDWI has been performed, in order to have a temporal trend of the used index and to assess the correlation between the fire ignition events and the status of the vegetation.

\section{RESULTS AND DISCUSSION}
\label{sec:results}

\begin{figure}[htb]
\begin{minipage}[b]{1.0\linewidth}
  \centering
  \centerline{\epsfig{figure=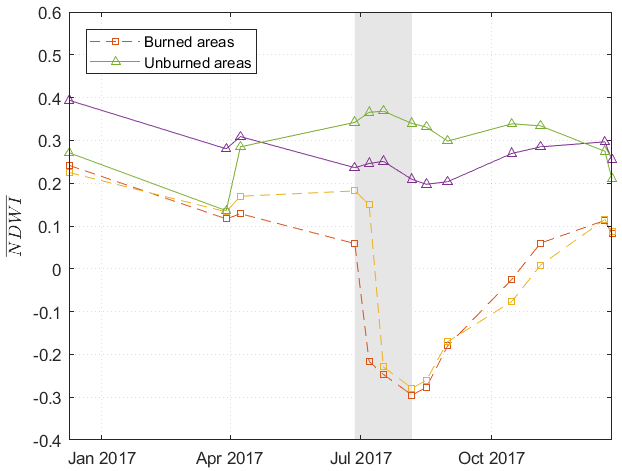,width=8.5cm}}
\end{minipage}
\caption{Time series of the NDWI mean on AOIs affected by fires (dashed lines) and on that ones were not (continuous line) for the Vesuvius area. Fires period is highlighted in grey.}
\label{fig:fig3}
\end{figure}

The results obtained by using the described method are shown in Fig.\ref{fig:fig3}.

Exploring the graphs we can notice that, after the fire event occurs (at the beginning of July 2017), the NDWI mean value in the burned areas rapidly decreases towards negative values. Let us remember that lower NDWI value indicates lower water content in the plants, and therefore the presence of a very dry vegetation. Even tough this appears an expected result, the fact that before the fire ignition event, in the area affected by fires the average NDWI is lower than that one of areas not strictly affected by fire is less expected.

\begin{figure}[!htb]
\begin{minipage}[b]{.48\linewidth}
  \centering
  \centerline{\epsfig{figure=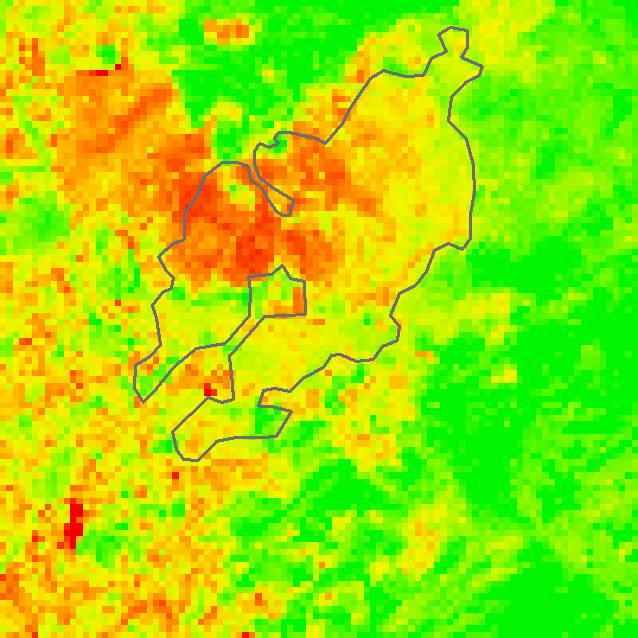,width=4.0cm}}
  \centerline{(a)}\medskip
\end{minipage}
\hfill
\begin{minipage}[b]{0.48\linewidth}
  \centering
  \centerline{\epsfig{figure=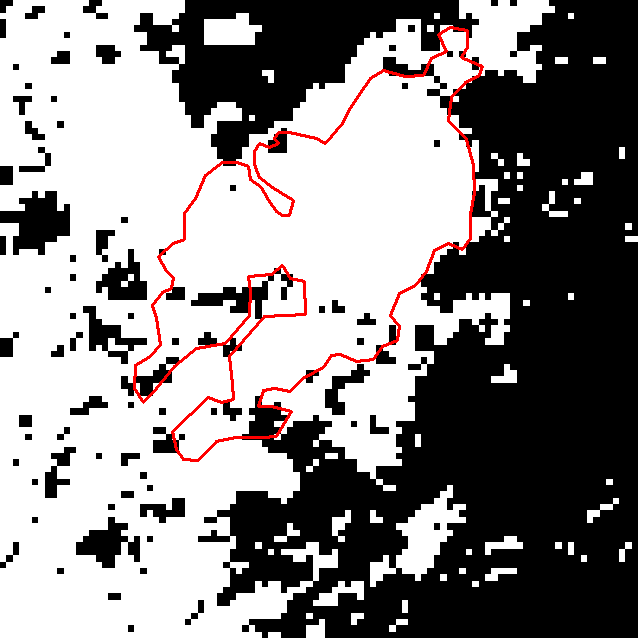,width=4.0cm}}
  \centerline{(b)}\medskip
\end{minipage}
\caption{NDWI map from the S-2 acquisition of the 26th June 2017 (a): from yellow to red the fire risk pixels. In (b), the binary map for the NDWI threshold $= 0.2$.}
\label{fig:fig4}
\end{figure}

Starting from this result, achieved performing the multi-temporal analysis, we investigated the possibility to produce a fire risk map based on the NDWI mean threshold, leading to a more clear and tangible view of this aspect. In fact, observing the NDWI mean values of the graph in Fig.\ref{fig:fig3}, we are able to separate the entire investigated area in two main classes: a burned-class, with the NDWI mean values less or equal than 0.2; and an unburned-class, with the NDWI mean values more then 0.2. Further confirmations on the validity of this threshold are presented in \cite{verhegghen2016potential}. The maps of the 26th June 2017 acquisition obtained using this threshold value are displayed in Fig.\ref{fig:fig4}, where a focus on a small area (67 hectares) of the volcano Vesuvius affected by the fire of 7th July 2017 is shown (i.e.,the outlined area).
    In order to visually validate this map, we need to define a Ground Truth (GT). For this work, such GT has been carried out by hand in GIS environment, using the information derived from the comparison between satellite and high resolution hyper-spectral images derived from a Visible-NIR pushbroom airborne sensor. Thanks to this GT it has been possible to discriminate on the Sentinel-2 data areas affected by fires from that ones which were not. 
As we can see in Fig.\ref{fig:fig4}, in the date before the fire event (26th June 2017), a significant part of the outlined area that was affected by fire presents NDWI values under the used threshold. Therefore, this value can be consider as a sort of \textit{risk threshold} based on the NDWI, for that area.

\section{CONCLUSIONS}
\label{sec:cocnlusion}

In conclusion, in this article we want to highlight how satellite remote sensing products can make a huge contribution to protecting the environment from the forest fires threat.

In particular, the use of the NDWI and the NDVI to monitor the water content of the vegetation areas, in order to perform a fires risk prevention task, has been tested. The encouraging results obtained by a multi-temporal analysis of the NDWI, followed by the confirmation derived from the visual inspection, allow us to further investigate and explore novel approach to develop useful remote sensing techniques for the ambitious forest fire prevention purpose. In particular, we can consider other indices and ancillary data as input, and deep-learning approaches that use a GT obtained by ground sensors and/or measurements.

\bibliographystyle{IEEEbib.bst}
\bibliography{refs.bib}

\end{document}